\documentclass[aps,prd,twocolumn,superscriptaddress,showpacs]{revtex4}
\usepackage{graphics}
\usepackage{epstopdf}
\usepackage{graphicx}
\usepackage{dcolumn}
\usepackage{bm}
\usepackage{amsmath}
\usepackage{color}
\usepackage{ulem}
\usepackage{lscape}
\usepackage{lineno,hyperref}
\usepackage[utf8]{inputenc}
\usepackage{amssymb}
\usepackage{footnote}

\usepackage{color}
\modulolinenumbers[5]
\usepackage{accents}
\newlength{\dhatheight}
\newcommand{\doublehat}[1]{%
    \settoheight{\dhatheight}{\ensuremath{\hat{#1}}}%
    \addtolength{\dhatheight}{-0.35ex}%
    \hat{\vphantom{\rule{1pt}{\dhatheight}}%
    \smash{\hat{#1}}}}

\newcommand{\be}{\begin{equation}}
\newcommand{\ee}{\end{equation}}
\newcommand{\beq}{\begin{eqnarray}}
\newcommand{\eeq}{\end{eqnarray}}

\def\t13{\mathrel{{\theta_{13}}}}
\def\y12{\mathrel{{\tan^2 \theta_{12}}}}
\def\c2{\mathrel{{\chi^2 }}}

\newcommand{\fl}{\textit{Fermi}-LAT}

\begin{document}


\title{Constraints on very high energy gamma-ray emission from the Fermi Bubbles with future  ground-based experiments}

\author{Lili Yang} \email{yanglli5@mail.sysu.edu.cn} 
\affiliation{School of Physics and Astronomy, Sun Yat-sen University, No.2 Daxue Rd, 519082, Zhuhai China}
\affiliation{Center for Astrophysics and Cosmology, University of Nova Gorica, Vipavska 11c, 5270 Ajdovscina, Slovenia}
\affiliation{Centre for Astro-Particle Physics (CAPP) and Department of Physics, University of Johannesburg, PO Box 524, Auckland Park 2006, South Africa}

\author{Soebur Razzaque} \email{srazzaque@uj.ac.za}
\affiliation{Centre for Astro-Particle Physics (CAPP) and Department of Physics, University of Johannesburg, PO Box 524, Auckland Park 2006, South Africa}


\begin{abstract} 
The origin of sub-TeV gamma rays detected by the \textit{Fermi}-Large Area Telescope (LAT) from the \textit{Fermi} Bubbles (FBs) at the Galactic center is still uncertain.  In a hadronic model, acceleration of protons and/or nuclei and their subsequent interactions with gas in the bubble volume can produce the observed gamma rays.  Recently the High Altitude Water Cherenkov (HAWC) observatory reported an absence of gamma-ray excess from the Northern FB at $b\gtrsim 6^\circ$ Galactic latitude, which resulted in flux upper limits in the energy range of $1.2-126$ TeV.  These upper limits are consistent with the gamma-ray spectrum measured by \textit{Fermi}-LAT at $|b|\ge 10^\circ$, where an exponential cutoff at energies $\gtrsim 100$ GeV is evident.  However, the FB gamma-ray spectrum at $|b|\le 10^\circ$, without showing any sign of cutoff up to around 1 TeV in the latest results, remains unconstrained. The upcoming Cherenkov Telescope Array (CTA) will perform a Galactic center survey with unprecedented sensitivity in the energy between 20 GeV and 300 TeV.  In this work, we perform both morphological and classic on/off analyses with planned CTA deep central and extended survey and estimate the sensitivity of CTA to the FB hadronic gamma-ray flux models that best fit the spectrum at $|b|\le 10^\circ$ and whose counterpart neutrino flux model best fits the optimistic neutrino spectrum from IceCube Neutrino Observatory.  We also perform sensitivity analysis with a future ground-based Cherenkov detector the Large High Altitude Air Shower Observatory (LHAASO). We find that CTA will be able to discover or constrain the FB gamma-ray flux at $|b|\le 10^\circ$ in the $\approx$ 200 GeV -- 100 TeV range with planned observation strategy, while LHAASO may constrain emission in the $\approx 100$~GeV -- 100~TeV range if $\lesssim 10\%$ systematic uncertainties can be achieved.  
\end{abstract}
 
\maketitle

\section{Introduction}
Ground-based Very High Energy (VHE, $\gtrsim 100$ GeV) gamma-ray astronomy has been undergoing dramatic progress over the past two decades, providing new insights into the known gamma-rays sources and developing strategies to search for potential new sources. The current imaging air-Cherenkov telescopes such as the High Energy Stereoscopic System (H.E.S.S.) \cite{Aharonian:2006pe}, Major Atmospheric Gamma-ray Imaging Cherenkov (MAGIC) telescopes \cite{Hassan:2015bwa} and Very Energetic Radiation Imaging Telescope Array System (VERITAS) \cite{Holder:2006gi}, and water-Cherenkov detector HAWC \cite{Abeysekara:2013tza} will be superseded by a new generation of observatories such as CTA \cite{Acharya:2017ttl} and LHAASO \cite{DiSciascio:2016rgi}. These upcoming experiments will achieve much better angular resolution and sensitivity, and will have wider field of view in comparison to the existing detectors.

Discovered in 2010 using \textit{Fermi}-LAT data \cite{Su:2010qj, Dobler:2009xz}, FBs at the Galactic center provide prime targets for the currently operational and upcoming VHE gamma-ray observatories.  Gamma-ray emission from the FB is spatially associated with Planck observation \cite{Ade:2012nxf}, the WMAP microwave haze \cite{Finkbeiner:2003im} and X-ray survey from ROSAT \cite{Mertsch:2011es}; confirming their multi-wavelength emission nature. These two bubbles have approximately uniform surface brightness in gamma rays at high latitude and have sharp rims, with an estimated luminosity of 4$\times 10^{37}$ erg/s between 100 MeV and 500 GeV \cite{Finkbeiner:2003im, Abeysekara:2017wzt}. 

The origin and production mechanism of gamma rays from the FB are still elusive. Both leptonic \cite{Su:2010qj, Mertsch:2011es, Fermi-LAT:2014sfa} and hadronic \cite{Crocker:2010dg, Crocker:2011en} models have been used to fit the gamma-ray spectrum successfully. In leptonic models, gamma rays are generated by inverse Compton scattering of infrared-ultraviolet photons by accelerated electrons. In hadronic models, gamma rays are produced by accelerated ions colliding with ambient particles in the bubble volume. High-energy neutrinos as counterparts to gamma rays are expected in case of a hadronic model \cite{Crocker:2010dg, Lunardini:2011br, Razzaque:2013uoa, Lunardini:2013gva, Lunardini:2015laa, Fermi-LAT:2014sfa, Fujita:2013jda}.  In the latest sample of 82 High-Energy Starting Events (HESE) with deposited energy above 60 TeV, released by the IceCube Collaboration \cite{Aartsen:2017mau}, the arrival directions of eight events (7 showers and 1 track events) are within the FB contours and the arrival directions of six shower events have their median positional error overlapping with the FB contours \cite{Razzaque:2018fzs}.  It is uncertain, however, how many of these neutrino events are from the FB itself or from a diffuse astrophysical background.  Another recent analysis shows up to $\approx 2\sigma$ probability that the FB is a neutrino source \cite{Fang:2017vlg}.  If confirmed with a higher significance in future larger HESE sample, FB might become the first diffuse source to have been seen in both gamma rays and neutrinos \cite{Lunardini:2015laa}.   

Recently the HAWC observatory presented their results on search for gamma-ray emission from the Northern bubble at Galactic latitude $b>6^\circ$ with 290 days of data \cite{Abeysekara:2017wzt}.  Without finding any significant excess, the HAWC observatory put constraints on the FB gamma-ray flux in the energy range of $1.2-126$ TeV. These flux upper limits are consistent with the latest FB gamma-ray data from the \textit{Fermi}-LAT at high-Galactic latitudes, $|b|\ge 10^\circ$ \cite{TheFermi-LAT:2017vmf}, which shows a cutoff in the spectrum at energies at $\gtrsim 100$ GeV.  The spectrum of the low-latitude ($|b|\le 10^\circ$) FB, however, is hard and does not show any obvious sign of cutoff at energies up to $\approx 1$ TeV.  The HAWC upper limits \cite{Abeysekara:2017wzt} therefore do not constrain the low-latitude FB spectrum, which could be a prime target for CTA and LHAASO in future. In addition, the spectrum in the central region of the FB may get harder at VHE range compared to the high-latitude spectrum, if there is a latitude-dependent emission. Therefore, we expect the central region of the FB would be an ideal target for studying the hadronic models, in particular to explore the VHE range. 

In this paper we estimate preliminary sensitivity of the CTA and LHAASO to the FB gamma-ray flux using publicly available information on the characteristics of these observatories. CTA South will be located in Paranal, Chile \cite{Acharya:2017ttl}, while CTA North will be located in La Palma, Canary Islands, Spain. There is already a strategy in place to survey the Galactic center region with CTA.  We exploit this survey characteristics in our calculation.  The LHAASO is currently under construction in Daocheng, China at 4410 m altitude \cite{DiSciascio:2016rgi} and will be able to observe the most part of the Northern bubble. 

The paper is organized as follows: we introduce the next-generation gamma ray observatories being constructed lately, CTA and LHAASO in Sec.\ II; we describe the hadronic models based on neutrino and gamma ray observations and discuss the major backgrounds in gamma-ray observation in Sec.\ III; we present our analysis methods, discuss the systematic errors for FBs, present our results and possible future constraints from gamma-ray observations in Sec. \ IV; we summarize and conclude in Sec.\ V.

\section{Ground-based gamma-ray observatories} 
Gamma-ray astronomy, focusing on the observation of photons with energies above 0.5 MeV and reaching up to a few hundred TeV, has achieved impressive progress over the last few decades. For the high-energy regime ($> 30$ MeV), gamma rays are either observed directly using spaced based telescopes or indirectly on the ground by detecting the electromagnetic showers produced by primary photons interacting with Earth's atmosphere. Currently, ground-based gamma-ray detectors are classified into two categories, namely the Imaging Atmospheric Cherenkov Telescope (IACT) and Extensive Air Shower (EAS) arrays based on the detection techniques. We refer to \cite{Thompson:2015nda, deNaurois:2015oda} for a review of gamma-ray detection techniques.  We discuss below briefly two upcoming detectors, namely CTA and LHAASO, in each of these categories and for which we estimate the sensitivities to the FBs. 

\subsection{The Cherenkov Telescope Array (CTA)} 
The IACTs observe gamma rays by detecting Cherenkov light emitted by initiated air shower of secondary particles in the atmosphere at high altitude.  Cherenkov light is beamed along the direction of primary particles and covers a circular area of about 250~m diameter on the ground.  Therefore, optical telescopes within the Cherenkov light pool will record images of the track of air shower. With multiple telescopes viewing the same primary particle, a stereoscopic reconstruction of the shower geometry is performed and precise information about the energy and arrival direction of the primary gamma rays are extracted. 
 
CTA is the fourth generation of IACTs for VHE gamma-ray astronomy and will be located in Northern and Southern Hemispheres for a full sky coverage.  It will cover a wide energy range from 20 GeV to up to 300 TeV \cite{Acharya:2017ttl}. CTA will improve the efficiency and sensitivity of current instruments by a factor of ten, resulting in excellent angular resolution and better discrimination of charged cosmic-ray backgrounds. According to its survey strategy, CTA will map the Galactic center region in unprecedented details, and will probe relationship between the central source and diffuse emission of the FBs for the first time in TeV energy range. Thus the maximum energy of the cosmic-ray accelerators responsible for gamma-ray emission and activity in the central 100~pc will be understood better. In addition, with its expected performance, CTA will possibly study the feature of sharp edges of the FBs above hundreds of GeV. 

\begin{figure}[htbp]
\centering
  \begin{minipage}[b]{1.0\linewidth}
    \centering
    \includegraphics[width=\linewidth]{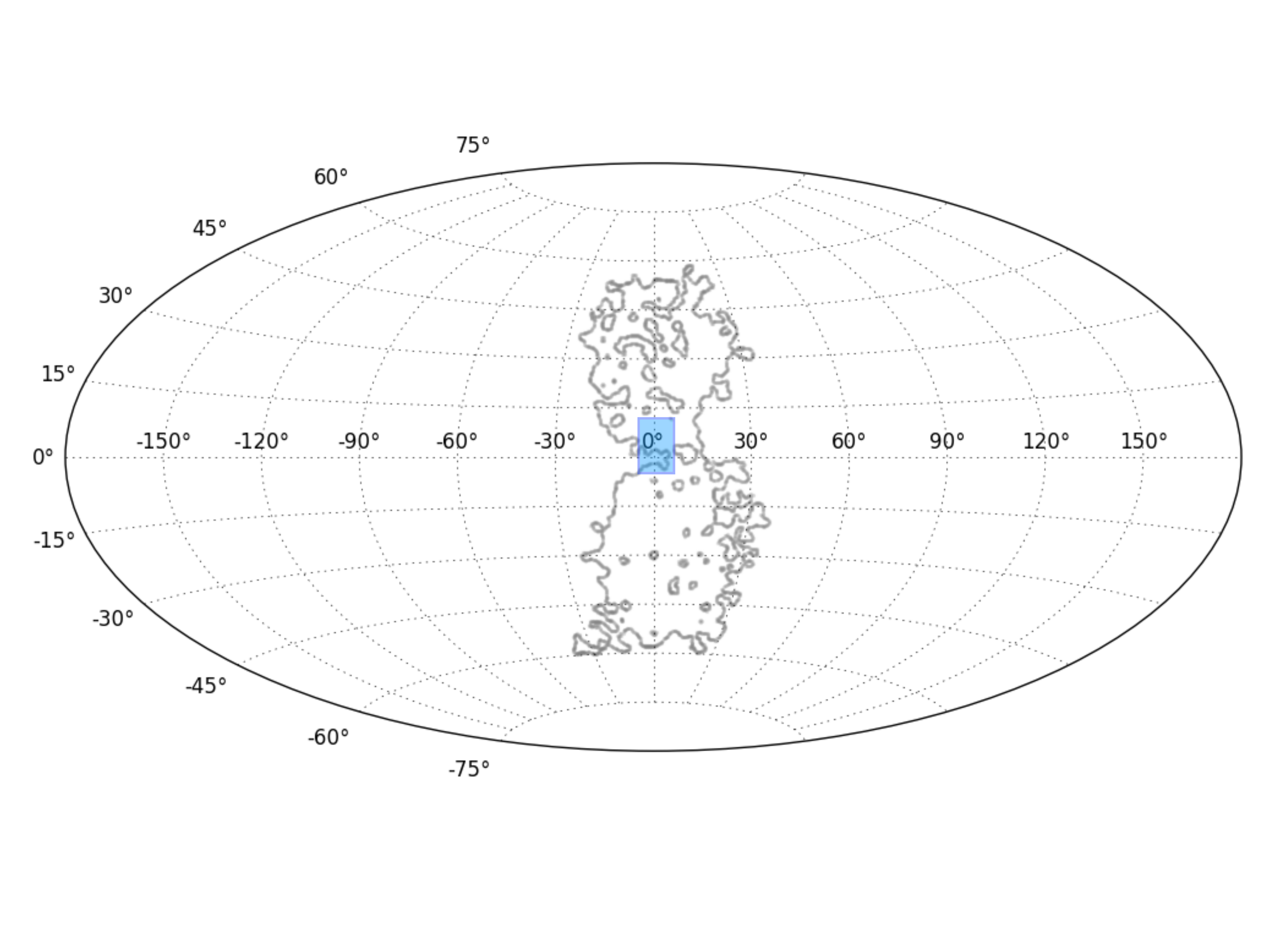}
  \end{minipage}
  \hspace{0.15cm}
  \begin{minipage}[b]{1.0\linewidth}
    \centering
    \includegraphics[width=\linewidth]{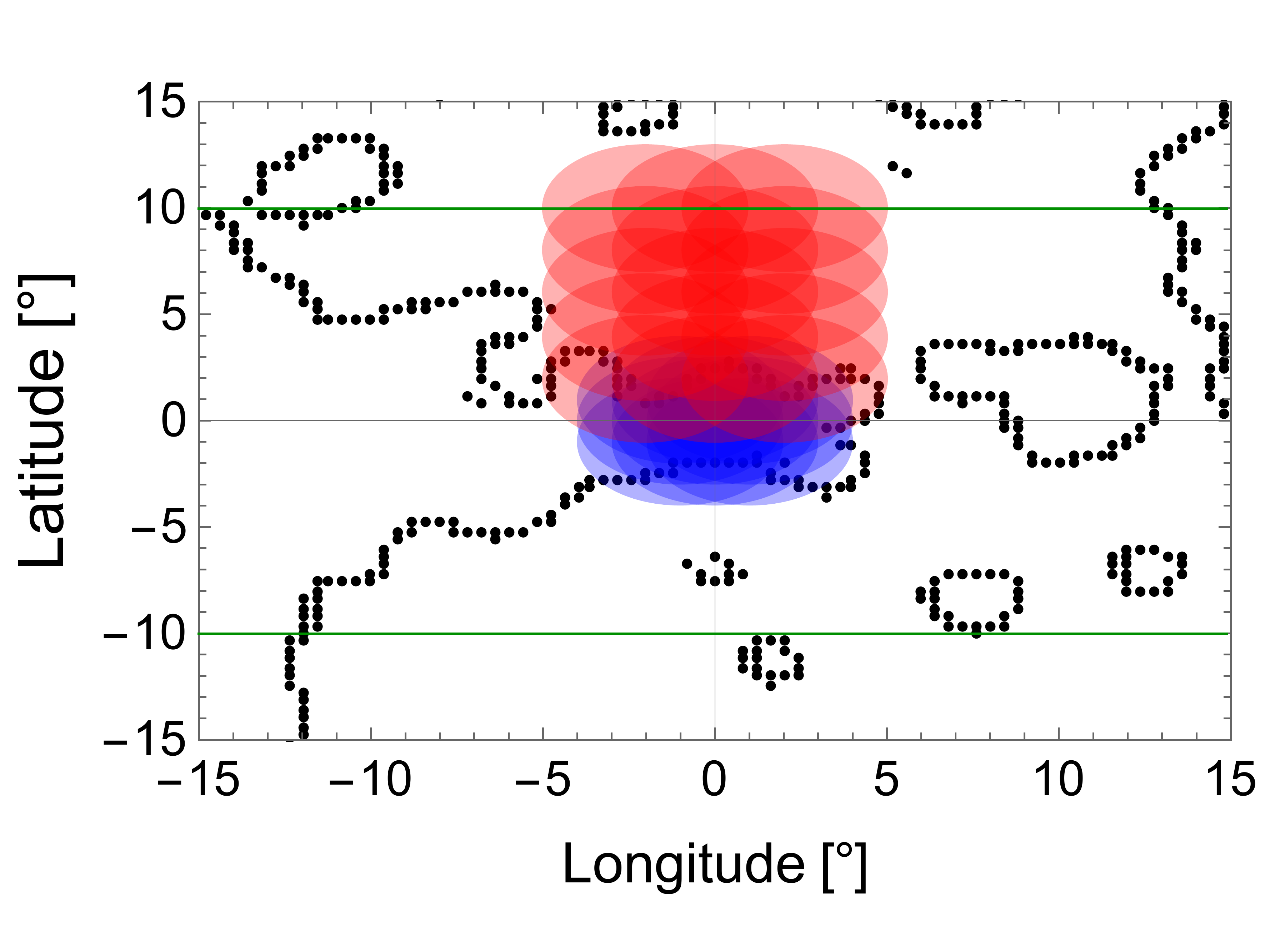}
  \end{minipage}
  \caption{Proposed CTA Galactic center and extended survey strategy in Galactic coordinate \cite{Acharya:2017ttl}. Upper panel shows the survey region as blue box and the \textit{Fermi} bubbles in gray contours. Lower panel presents the central 9-point scanning (525 hour) as blue circles, extended 15-point scanning (300 hour) with red circles, and the \textit{Fermi} bubbles in black dotted contours. For each pointing, the radius is $3^\circ$. }
\label{fig:skymap}
\end{figure}

In this study, we will concentrate on study of the FBs with the CTA Southern array based on current configuration, referred to as \texttt{Production 3b} \cite{Hassan:2015bwa, Acharya:2017ttl}. To better target the FB central region, we perform our study according to the CTA proposed pointing strategy \cite{Acharya:2017ttl} of a deep exposure centered on the Sgr A* and extended region survey as illustrated in Fig.~\ref{fig:skymap}. In the first year's operation, the central survey will perform 525 hours of observation centered on the Galactic center with 9 pointing (at 0$^\circ$ and $\pm$ 1$^\circ$ in $l$ and $b$); the extended survey will have 300 hours of observation on the Northern Galactic plane with 15 pointing (at 0$^\circ$ and $\pm$ 2$^\circ$ in $l$, 2$^\circ$ to 10$^\circ$ in $b$). We apply \texttt{ctools (v1.5.0)} \cite{Knodlseder:2016nnv}\footnote{http://cta.irap.omp.eu/ctools/}, a dedicated software package for scientific analysis of CTA, to simulate the observation, where a radius of 3$^\circ$ field of view for each pointing is adopted. Even though FB detection is very challenging, due to its large structure and complicated emission from the inner part ($|b| < 10^{\circ}$), our work will give the first sensitivity estimation for this giant diffuse source with stacking analysis.

\subsection{The Large High Altitude Air Shower Observatory (LHAASO)}
The extended air shower (EAS) gamma-ray detectors such as Pierre Auger and HAWC have proven to be successful. They have high duty cycle and large field of view. The LHAASO project will be one of the world's largest cosmic-ray observatories, currently being constructed at Haizi Mountain in Sichuan province, China \cite{DiSciascio:2016rgi}.  It aims at detecting both gamma rays and cosmic rays over wide energy ranges from $10^{11}$~eV to $10^{15}$~eV and from $10^{12}$~eV to $10^{17}$~eV, respectively.  LHAASO will consist a water Cherenkov detector array (WCDA), 12 wide-field fluorescence telescopes above ground, a 1.3 km$^2$ array (KM2A) made of 6000 plastic scintillators and approximately 1146 Cherenkov water tanks spread over the same area underground.  In the energy range of interest to search for gamma rays from the FBs, one should utilize the WCDA for FB detection. 

According to LHAASO's location, it will be able to observe the most part of the Northern FB. Until now HAWC does not provide sensitivity in the sub-TeV energy regime, due to large signal contamination and background, and their work to improve sensitivities is still going on. However, the LHAASO surface water array will be four times larger than HAWC and therefore will be more sensitive. As complementary to CTA's sensitivity to the FBs, LHAASO will provide additional information of the central regions of the FB within a shorter time scale and at higher energies.  

\section{Emission from the Fermi Bubbles and backgrounds}
\begin{figure}[htb]
\centering
\begin{minipage}[b]{0.98\linewidth}
\centering
\includegraphics[width=\linewidth]{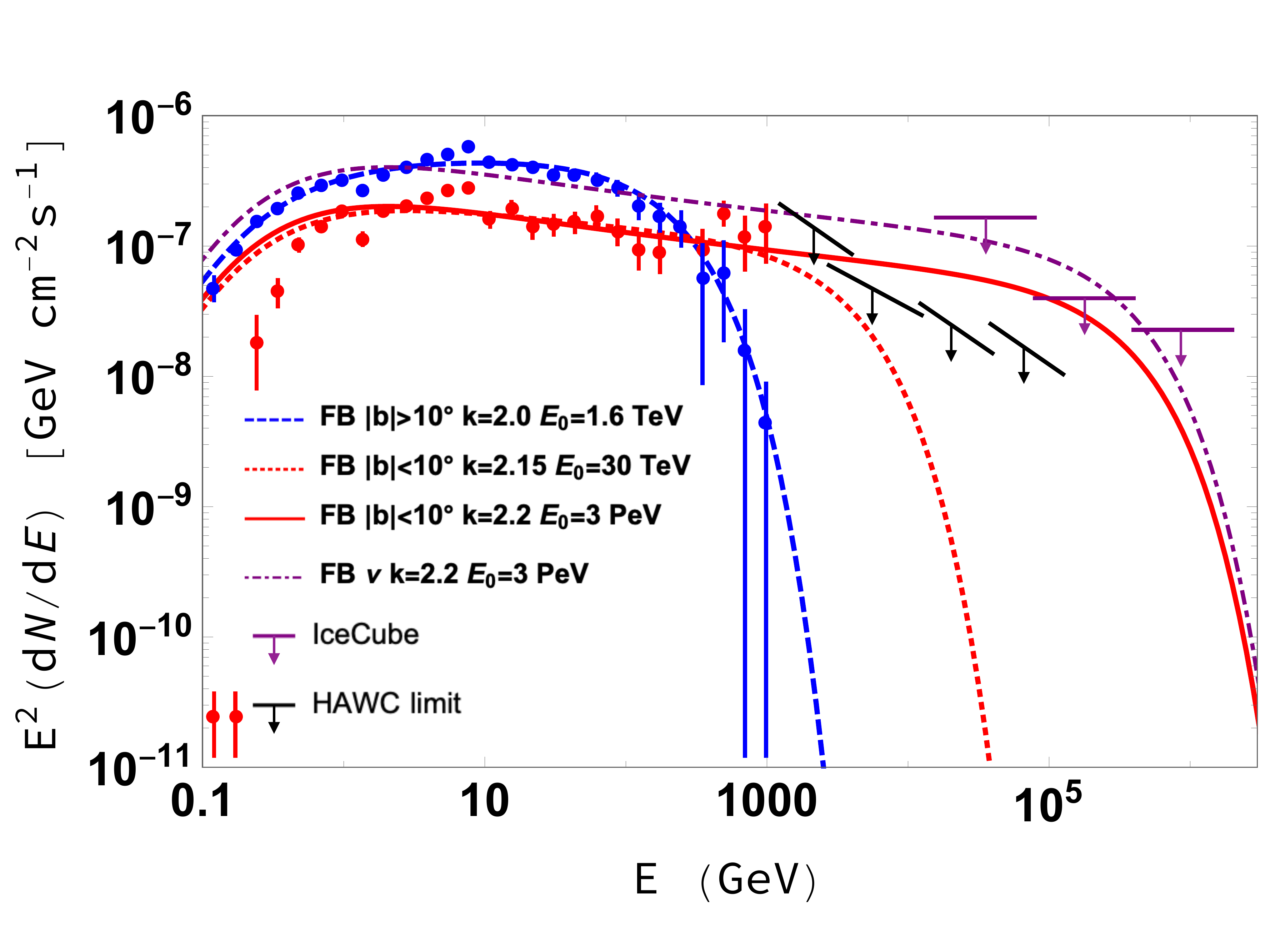}
\end{minipage}
\caption{Gamma-ray spectra from the FB at the low-latitude and high-latitude regions~\cite{TheFermi-LAT:2017vmf}. The HAWC upper limits are derived in the $b>6^\circ$ region~\cite{Abeysekara:2017wzt}. The neutrino upper limits are calculated from IceCube events spatially correlated with the FB~\cite{Aartsen:2017mau}. The hadronic models are shown with primary proton spectral index $k$ and exponential cutoff energy $E_0$ combination for $(k, E_0) = $ (2.0, 1.6~TeV) blue dashed line; (2.15, 30~TeV) red dotted line; and (2.2, 3~PeV) red solid lines. The magenta dashed line shows the neutrino flux (all three flavors) for the model with $(k, E_0) =$ (2.2, 3~PeV). Figure adapted from Ref.~\cite{Razzaque:2018fzs}.}
\label{fig:spectra}
\end{figure}

In this work we have adopted the FB template and spectra from recent Galactic center analysis of \textit{Fermi}-LAT using 6.5 years of data with the region of interests in longitude $|b| < 60^{\circ}$ and latitude $|l| < 45^{\circ}$ \cite{TheFermi-LAT:2017vmf}. The analysis was done with spectral components analysis (SCA) procedure as in Ref.~\cite{Fermi-LAT:2014sfa}, where the residuals were obtained by subtracting the gas-correlated emission and point source models from the data and then were decomposed into two components correlated with $E^{-2.4}$ and $E^{-1.9}$ spectra. In the end, the FB template was derived by cutting the hard component $E^{-1.9}$ at $2\sigma$ significance. Notice that in this updated FB analysis the spectrum of the low-latitude ($|b|<10^\circ$) FB is similar to the spectrum of the high-latitude ($|b|>10^\circ$) FB at lower energy from 100 MeV to 100 GeV, but remains hard above 100 GeV as can be seen in Fig.~\ref{fig:spectra}.  Also shown in Fig.~\ref{fig:spectra} the HAWC upper limits in the $b>6^\circ$ region \cite{Abeysekara:2017wzt}.

\subsection{Hadronic model}
We adopt the same hadronic models as in \cite{Lunardini:2011br, Lunardini:2015laa} but updated in Ref.~\cite{Razzaque:2018fzs} with latest \textit{Fermi}-LAT ~\cite{TheFermi-LAT:2017vmf} and IceCube data~\cite{Aartsen:2017mau}. We assume an exponentially cut-off power law proton injection spectrum in the form of $\sim$ $E^{-k} \exp(-E/E_0)$, where $k$ and $E_0$ are the spectral index and exponential cutoff energy, respectively.  The blue dashed ($k=2.0$, $E_0 = 1.6$~TeV) and red dotted ($k=2.15$, $E_0 = 30$~TeV) lines are the best-fit hadronic models for the low-latitude and high-latitude FB emission regions, respectively. The solid red line ($k=2.2$, $E_0 = 3$~PeV) which is very similar to the red dashed line at and below the highest LAT data point, is inspired by requiring neutrino flux from the same hadronic model to satisfy the upper limits calculated from 8 IceCube events in 6-year IceCube neutrino data \cite{Aartsen:2017mau} from the direction of the FB originating from the hadronic model \cite{Razzaque:2018fzs}. Note that our best-fit gamma-ray flux models are compatible with the HAWC upper limits.  The neutrino inspired model gives gamma-ray flux higher than the HAWC upper limits at $\gtrsim 3$~TeV but these upper limits do not strictly apply to the low-latitude FB flux. We refer readers to our recent review~\cite{Razzaque:2018fzs} for more details. 

                                                
\subsection{Background models}
\begin{figure}[htbp]
\centering
  \begin{minipage}[b]{1.0\linewidth}
    \centering
    \includegraphics[width=\linewidth]{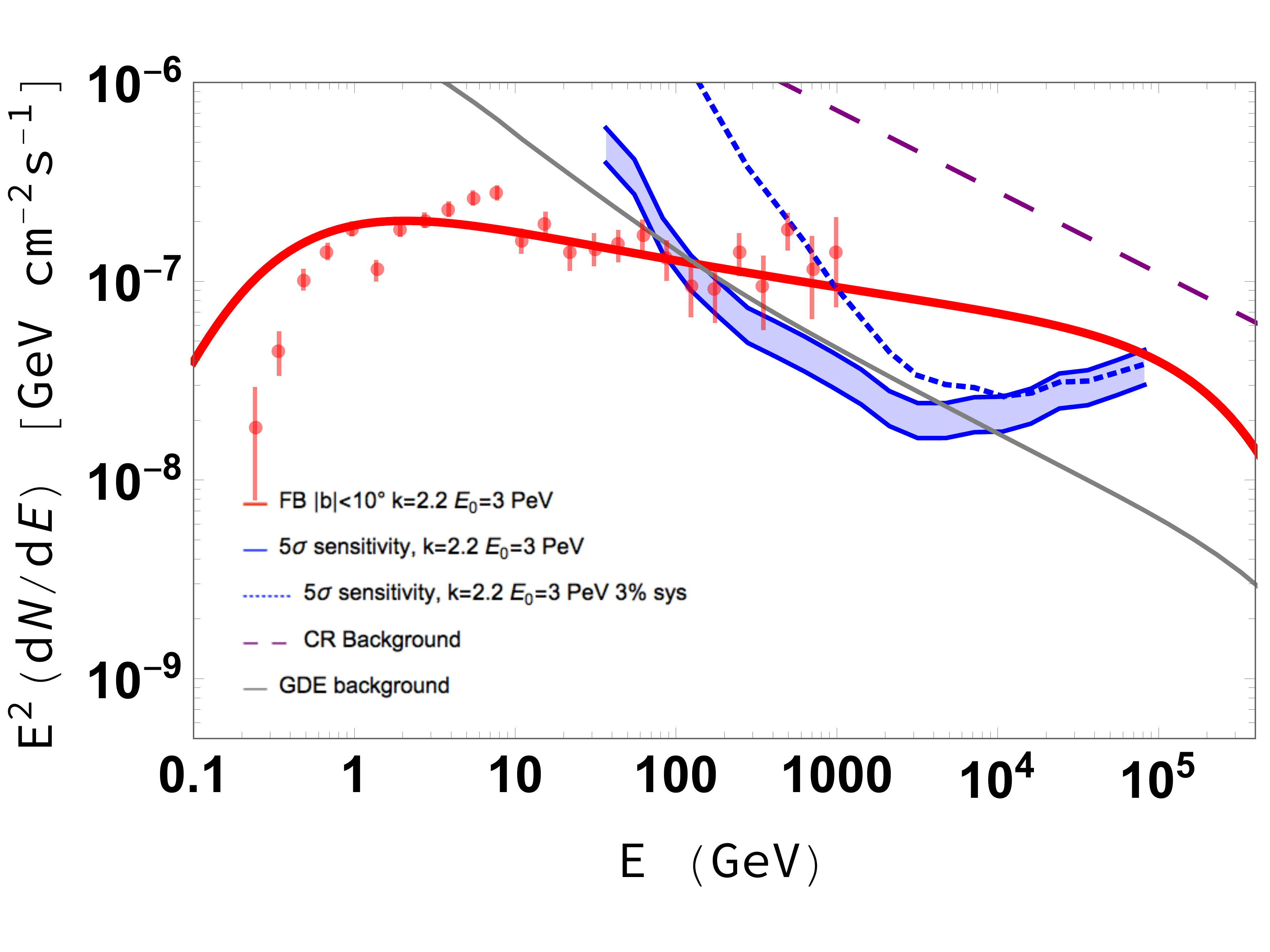}
  \end{minipage}
  \hspace{0.15cm}
  \begin{minipage}[b]{1.0\linewidth}
    \centering
    \includegraphics[width=\linewidth]{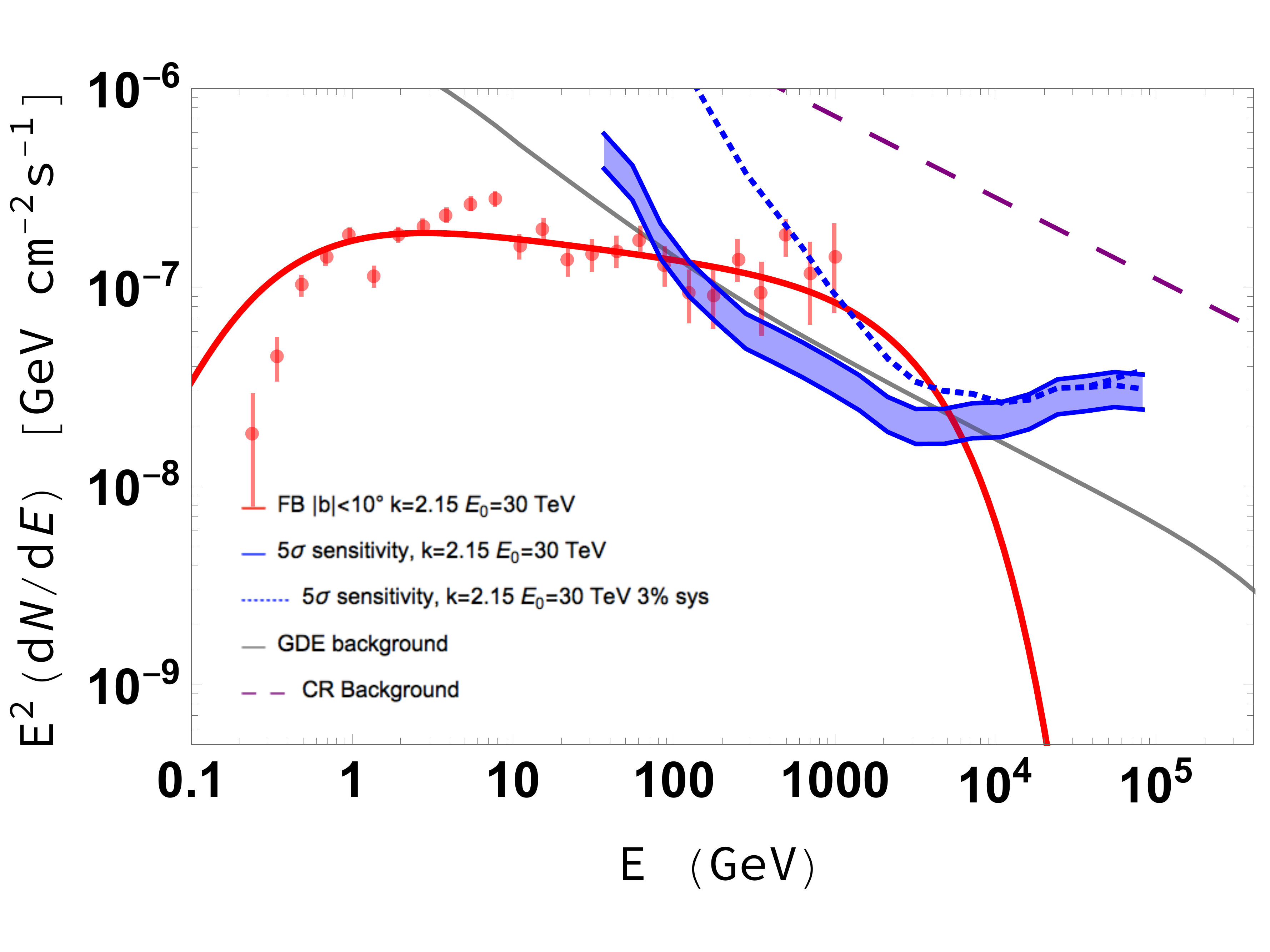}
  \end{minipage}
    \hspace{0.15cm}
  \begin{minipage}[b]{1.0\linewidth}
    \centering
    \includegraphics[width=\linewidth]{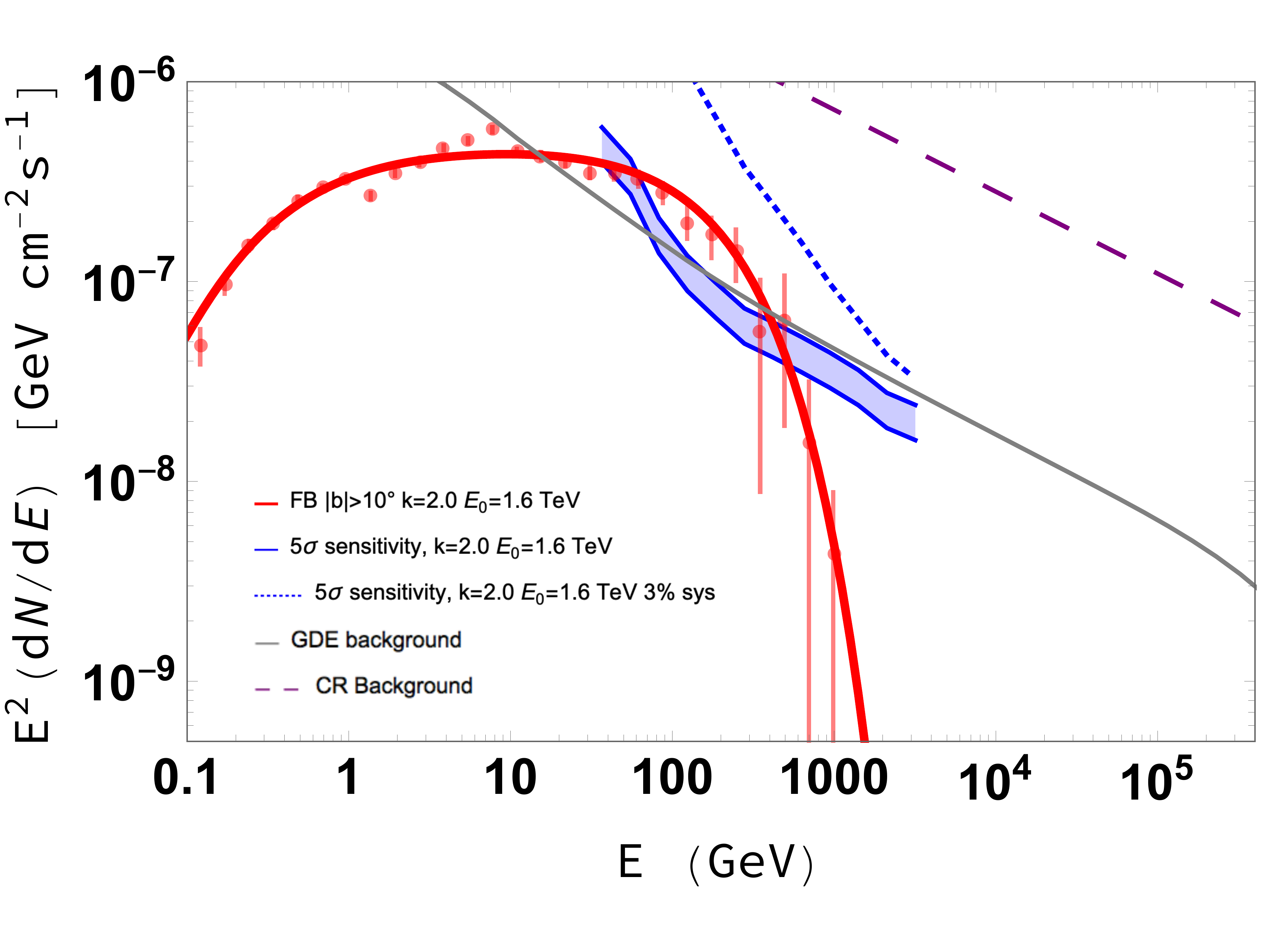}
  \end{minipage}
  \caption{CTA sensitivity to hadronic models for the FB gamma-ray spectra (red curves). Top panel is for the high-cutoff spectral model ($k=2.2$, $E_0=3$ PeV), middle panel is for the low-cutoff spectral model ($k=2.15$, $E_0=30$ TeV) and bottom panel is the model at high attitude ($k=2.0$, $E_0=1.6$ TeV) respectively. The CR and GDE backgrounds are shown with purple long-dashed and gray solid lines respectively. The red points in upper and middle (bottom) panels are data from the low- (high-)latitude region of Fermi observation~\cite{TheFermi-LAT:2017vmf}. The CTA sensitivities with statistical uncertainty to the hadronic models in the low-latitude region are shown in blue shaded region in case of no systematic uncertainty and shown as blue dotted curves with 3\% systematic uncertainty.  
}
\label{fig:FB_cta}  
\end{figure}

For IACTs, the major source of background is cosmic rays (CRs), namely hadrons and electrons, whose flux is about three orders of magnitude higher than that of gamma-ray signals. The background rejection requires efficient techniques, such as identification of shower image and distribution of arrival time of the shower front \cite{Voelk:2008fw}. In addition, thanks to the application of machine-learning algorithms, the background rejection efficiency has been further improved. Even though few CRs, which are mostly electrons, can trigger and pass the photon cuts, the residual contamination is still expected. In this work, we adopt the instrumental background implemented in ctools, which is derived through extensive Monte Carlo simulation for the planned array layout \texttt{prod 3b}. Therefore, the dominant backgrounds for studying the FB are residual CRs and diffuse gamma-ray emission from the Galactic center region.  At CTA, the residual CR background is simulated and analyzed based on the instrumental properties, such as the effective area and point spread function. The cosmic-ray background of Instrument Response Function (IRF) follows a power law of shape $\sim$ $E^{-2.41}$ as shown with purple long-dashed lines in Fig.~\ref{fig:FB_cta}. The second dominant background is the Galactic Diffuse Emission (GDE) which has been observed by \textit{Fermi}-LAT~\cite{Atwood:2013rka} and H.E.S.S.~\cite{Aharonian:2006au}. We have adopted a GDE model (gray solid lines in Fig.~\ref{fig:FB_cta}) following the study of gamma-ray emission in Ref.~\cite{Gaggero:2017jts}, which interpret the current data with a CR transport model and is consistent with the contribution of $\pi^{0}$ decay, inverse Compton and bremsstrahlung emissions. Other contributions to the gamma-ray background are point and extended sources located at or near the Galactic center. On the other hand, main backgrounds for LHAASO are hadronic cosmic rays from proton to iron, whose combined flux is $\sim 3$--4 orders of magnitude larger than the flux of FB gamma rays.  Since the cosmic electron spectrum is about two orders of magnitudes lower than the flux of hadrons, its contribution to the background is neglected in our analysis hereafter.

\section{Analysis}
Both CTA and LHAASO are under construction, therefore there is no data available to test the FB flux models yet.  Instead we use the well-known Asimov data set \cite{Cowan:2010js} constructed from the flux models and corresponding expectation values to save the computing power required for simulations, since it has been well justified and widely applied to estimate the median experimental significance \cite{Cowan:2010js}.  We perform both a morphological analysis and an ON/OFF analysis for CTA using the Asimov data set in order to calculate sensitivity to the FB flux models and identify sharp edge of the FB.  For LHAASO, we use a simplified method to estimate sensitivity to the FB using signal and background events.  The details are discussed below.     

\subsection{Morphological analysis for CTA}
The dataset is binned with 0.5 degree per pixel and 20 energy bins in the energy range from 30 GeV to 100 TeV.  To calculate the CTA sensitivity to the FB, we adopt the method of statistical hypothesis test.  The probability of observing $n$ events with $m$ expected events is given by the binned Poisson likelihood function 
\begin{equation}
\mathcal{L}_i =\prod_j \frac{m_{ij}^{n_{ij}}e^{-m_{ij}}}{n_{ij}!},
\label{eq:likelihoodfun}
\end{equation}
where $i$ and $j$ represent $i$-th energy bin and $j$-th region of interest. The model data $m_{ij}$ can be expressed with three components as $m_{ij}=\beta_{i,1} b_{ij,\rm CR} + \beta_{i,2} b_{ij,\rm GDE} + \mu_{i} s_{ij,\rm FB}$. Here $b_{ij,\rm CR}$, $b_{ij,\rm GDE}$ and $s_{ij,\rm FB}$ are the expected numbers of events in each pixel generated by \texttt{ctmodel} of ctools based on instrumental CR background, GDE model and derived FB model, and $\mu_i$, $\beta_{i,1}$ and $\beta_{i,2}$ are the strength parameter and nuisance parameters in each energy bin. For including the point sources in the analysis, such as central gamma-ray source HESS J1745--290 \cite{Aharonian:2004wa}, the pulsar wind nebula G0.9+0.1 \cite{Aharonian:2005br} and supernova remnant (SNR) HESS J1745--303 \cite{Aharonian:2008gw}, one needs to include additional components in the formula here. Including the background point sources in our analysis however does not affect our results. This is because the point sources contribute only a few events in 4 to 6 pixels over a total of 10$\times$10 pixel region at each energy bin. Therefore, the maximum likelihood (ML) ratio can be defined as \cite{Cowan:2010js}
\begin{equation}
\lambda_i = \frac{\mathcal{L}_i(n|m(\mu_i,\doublehat{\beta_{i,1}},\doublehat{\beta_{i,2}}))}{\mathcal{L}_i(n|m_0(\widehat{\mu_i},\widehat{\beta_{i,1}},\widehat{\beta_{i,2}}))} .
\label{eq:lratio}
\end{equation}
In Eq.~(\ref{eq:lratio}), the denominator is the unconditional ML with $\widehat{\mu}$, $\widehat{\beta_1}$ and $\widehat{\beta_2}$ as ML estimators fitted to the Asimov data set.  The numerator is the conditional ML, which is maximized with $\doublehat{\beta_1}$ and $\doublehat{\beta_2}$ for specified value of $\mu$.  Therefore the ratio $\lambda$ in Eq.~(\ref{eq:lratio}) is related to $\chi^2$ with 1 degree of freedom (the difference in degrees of freedom between two hypothesis) as $2~\text{ln} \lambda \sim \chi_1^2$, based on Wilks' theorem \cite{Wilks:1938dza}. To compute differential CTA sensitivity, the mock data in each pixel is the total number of background events without the test source (FB), $n_{ij}=b_{ij,\rm CR} + b_{ij,\rm GDE}$. As the discovery significance of a detection is simply estimated as the square root of the Test Statistic (TS) value, where TS = $-2~\text{ln} \lambda$. We obtain the strength factor $\mu_i$ by setting $- 2~\text{ln} \lambda_i = 25$ corresponding to 5 sigma significance. Therefore, the sensitivity is computed as $\mu_i\times E_{0,i}^2\times\phi(E_{0,i})$, where $E_{0,i}$ is logarithmic mean energy of the \textit{i}-th energy bin, and $\phi(E_{0,i})$ is the corresponding FB flux density. To take into account statistical fluctuation of the CTA discovery significance with Asimov data set, we follow Eq.~(31) in Ref.~\cite{Cowan:2010js} to obtain the standard variance $\sigma_{\mu i}$ of $\mu_i$.  Therefore sensitivities with statistical uncertainty are shown as the shaded region in Fig.~\ref{fig:FB_cta}, corresponding to $\mu_i \pm \sigma_{\mu i}$.  For minimization of the logarithm likelihood function, we utilize the python \texttt{iMinuit} algorithm \cite{James:1975dr}\footnote{https://github.com/iminuit/iminuit} and find the best fit parameters. 

Major systematic uncertainties in our study are from instrumental response. As the number of background CRs is much more than signal photons, the instrumental systematic uncertainty is the primary concern in the analysis, even though the systematic uncertainty from GDE models can get up to 10\% -- 20\%. Following \cite{Silverwood:2014yza}, we multiply the model data $m_{ij}$ by a scaling parameter $\theta_{ij}$ to account for the independent pixel-to-pixel systematic effects in the observation.  We assume that the nuisance parameters follow a Gaussian distribution with variance $\sigma$ for all $\theta_{ij}$, independent of $i$ and $j$. Therefore the likelihood function in Eq.~(\ref{eq:likelihoodfun}) can be written as \cite{Silverwood:2014yza}.
\begin{equation}
\mathcal{L}_i(n | m, \theta)=\prod_j \frac{(\theta_{ij} m_{ij})^{n_{ij}}}{\sqrt{2\pi}\sigma n_{ij}!}e^{-\theta_{ij} m_{ij}}e^{-\frac{(1-\theta_{ij})^2}{2\sigma^2}} \,.
\label{eq:like1}
\end{equation}
The ML value of $\theta_{ij}$ for each $i$ and $j$ can be obtained by solving the derivative of Eq.~(\ref{eq:like1}) with respect to $\theta_{ij}$ set to zero as \cite{Silverwood:2014yza}.
\begin{equation}
\theta_{ij}=\frac{1}{2}\Big(1-\sigma^2m_{ij}+\sqrt{1-2\sigma^2m_{ij}+4\sigma^2n_{ij}+\sigma^4m_{ij}^2}\Big) \,.
\label{eq:sys}
\end{equation}

For IACTs such as H.E.S.S., the acceptance inhomogeneity is estimated to be 3\% or less, unless observations is done with large zenith angle or bright sources in the FoV \cite{Berge:2006ae}. Therefore, we take the value of variance $\sigma = 0.03$ as the case for our analysis. The 5 sigma sensitivity for the case with systematic uncertainty are shown as dotted blue curve in Fig.~\ref{fig:FB_cta}.  The top (bottom) panel shows these sensitivities for the high- (low-) spectral cutoff model for the low-latitude FB flux. The obtained sensitivities clearly show that CTA will have good sensitivity to the FB flux models in the $\gtrsim 70$ GeV range, overlapping with \textit{Fermi}-LAT data. Also, as seen in Fig. \ref{fig:FB_cta}, systematics are the major effect at lower energy ($<$ 2 TeV), however at higher energy ($>$ 2 TeV), statistics take more role in the signal discovery. This is because at lower energy the background is more dominant, whose detection is strongly dependent on the instrumental systematics and at high energy the sensitivity is limited by a low signal flux. As seen in the bottom panel of Fig.~\ref{fig:FB_cta}, when including the 3\% systematic uncertainty, it will be difficult for CTA to observe the high-latitude region of the FB. Further observational strategies will need to be considered in this case.

\subsection{On/Off analysis for CTA}
\begin{figure}[htbp]
\vskip 0.4 cm
\centering
  \begin{minipage}[b]{1.0\linewidth}
    \centering
    \includegraphics[width=\linewidth]{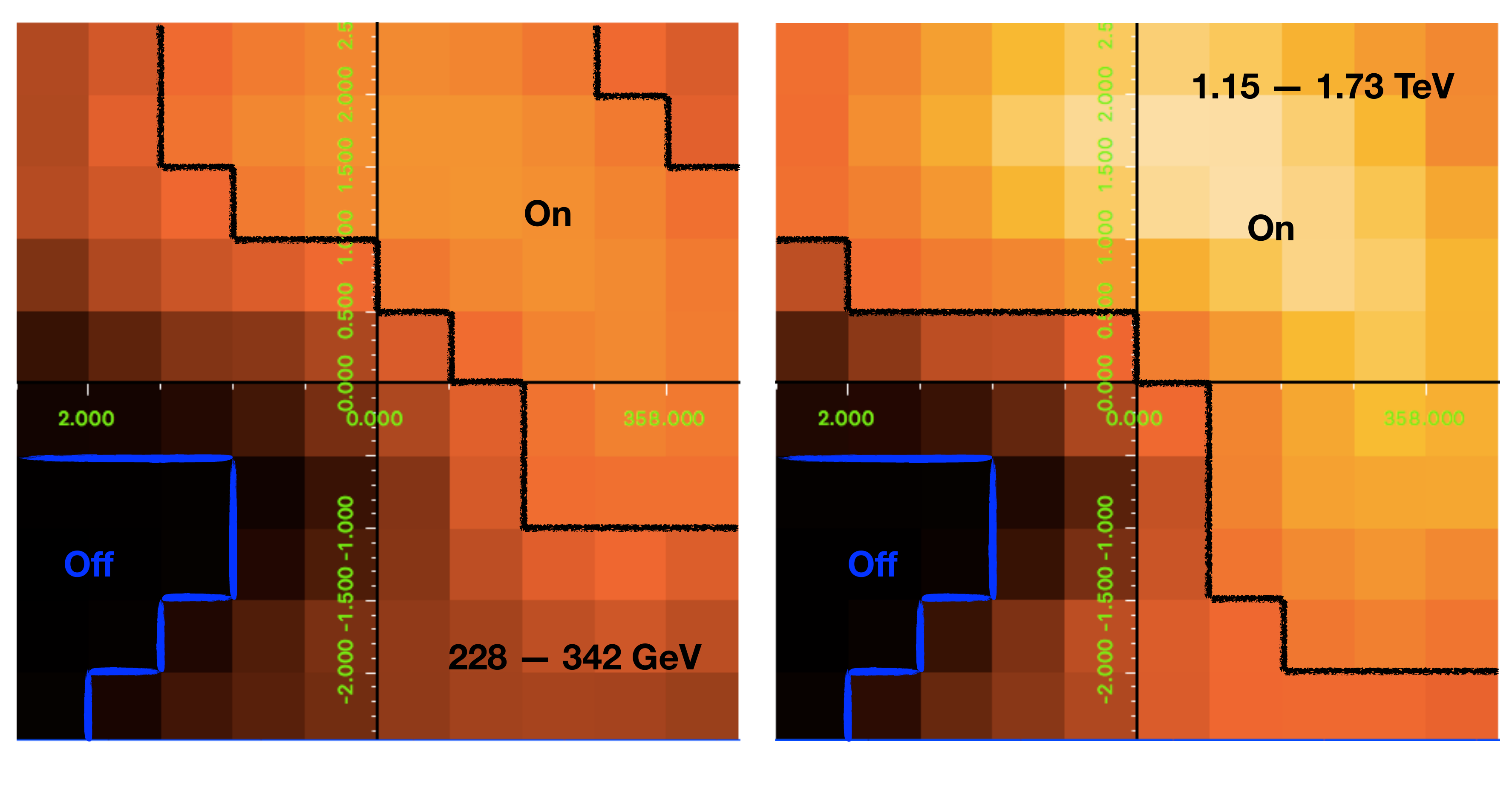}
  \end{minipage}
  \hspace{0.15cm}
  \begin{minipage}[b]{1.0\linewidth}
    \centering
    \includegraphics[width=\linewidth]{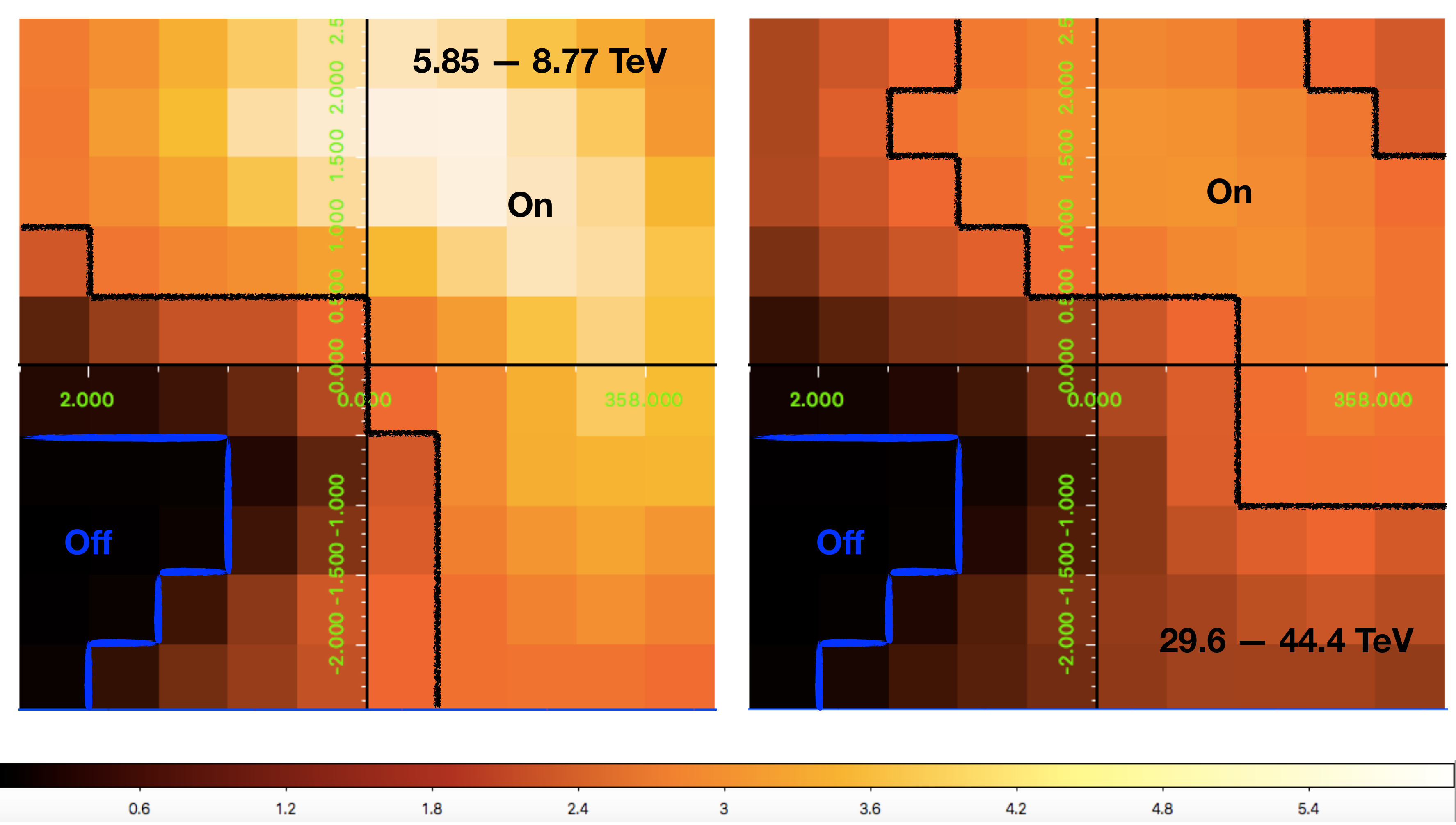}
  \end{minipage}
  \caption{Calculated significance $\Delta_{ij}$ to the FB with CTA in each pixel of four energy bins, 228 -- 342 GeV (top-left), 1.15 -- 1.73 TeV (top-right), 5.85 -- 8.77 TeV (bottom-left) and 29.6 -- 44.4 TeV (bottom-right). The selected ON and OFF regions are encircled with black and blue lines, respectively. High spectral cutoff model ($k=2.2$ $E_0 = 3$ PeV) has been used here for the low-latitude FB spectrum.}
\label{fig:cta_sig}  
\end{figure}

In a classic ON/OFF analysis or the so called Li-Ma analysis \cite{Li:1983fv}, it is interesting and ideal to use FB edges as border between the ON and OFF regions to test if we can find sharp rims in the observation. First of all, we calculate the statistical significance $\Delta_{ij}$ in each pixel, $\Delta_{ij}=\sqrt{2(n_{ij}\text{ln}(1+s_{ij}/b_{ij})-s_{ij})}$ for the Asimov dataset with known backgrounds.  As for next step, we select the ON pixels with certain gamma-ray excess $\Delta_{ij} > 2$ and the OFF pixels with no significant gamma-ray excess, $\Delta_{ij} \leq 0.1$.  Statistical significance of the pixels is shown in Fig.~\ref{fig:cta_sig} together with selections of the ON/OFF regions for the high spectral cutoff model ($k=2.2$ $E_0 = 3$ PeV), where we can easily identify the regions with apparent brightness. However, the estimated significance explicitly depends on background and GDE models, which lack sufficient data at TeV regime and which need to be improved and modified according to future measurement.  Systematic uncertainty from the GDE model has also been taken into account as discussed in Sec.\ IV A. For the high-cutoff FB spectrum, there is no pixel passing the ON selection for the first four and the last energy bins (30 -- 152 GeV and 67 -- 100 TeV, respectively), where the background events dominate over signal events. Hence, the energy range used for this analysis is from 228 GeV to 100 TeV. For the low spectral cutoff model ($k=2.15$ $E_0 = 30$ TeV), half of the energy bins have the selected ON regions, ranging from 101 GeV to 5.85 TeV.  As a result, we exclude those bins without the ON-region in the calculation.

\begin{figure}[h!]
\begin{centering}
\includegraphics[width=3.4truein]{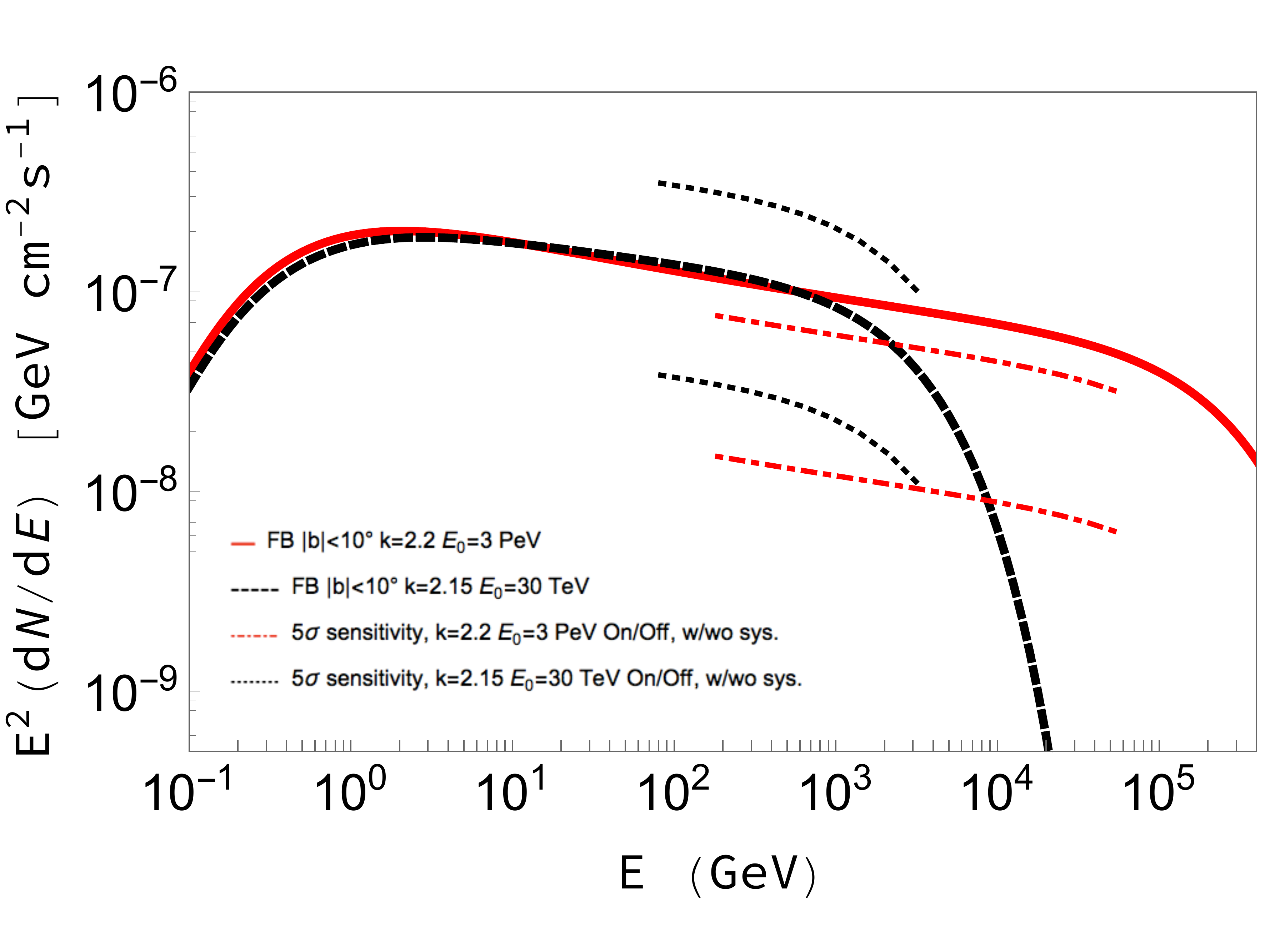}
\vskip -0.05truein
\caption{CTA sensitivity for detecting FBs using ON/OFF analysis for two hadronic gamma-ray spectra for the low-latitude region, namely the high-cutoff ($k=2.2$, $E_0=3$ PeV) and low-cutoff ($k=2.15$, $E_0=30$ TeV) models, are shown as red dot-dashed and black dotted curves, where the lower (upper) curve is for the case without (with) 3\% systematics respectively. }
\label{fig:cta_onoff}   
\end{centering}
\end{figure}

For the Asimov dataset with known backgrounds, the parameter $\tau$ is denoted as the ratio of the expected means of $b_{\rm off}$ and $b_{\rm on}$, i.e., $\tau=b_{\text{off}}/b_{\text{on}}$.  The corresponding likelihood function for the ON/OFF problem can be expressed as  in \cite{Cousins:2007bmb} 
%
%
\begin{equation}
\begin{split}
\mathcal{L_{\text{on/off}}}&= \prod_i \frac{ m_{\text{on},i}(\mu,\beta_1,\beta_2)^{n_{\text{on},i}}}{n_{\text{on},i}!}e^{-m_{\text{on},i}(\mu,\beta_1,\beta_2)}\\
             &\times \frac{m_{\text{off},i}(\beta_1,\beta_2,\tau)^{n_{\text{off},i}}}{n_{\text{off},i}!}e^{-m_{\text{off},i}(\beta_1,\beta_2,\tau)} \,,
\end{split}
\end{equation}
%
%
where $m_{\text{on}}$ and $m_{\text{off}}$ are expected number of events for the ON and OFF region, given as $m_{\text{on},i}=\sum_j \big(\beta_1 b_{ij,\rm CR} + \beta_2 b_{ij,\rm GDE} + \mu s_{ij,\rm FB}$\big) and $m_{\text{off},i}=\tau \sum_j \big(\beta_1 b_{ij,\rm CR} + \beta_2 b_{ij,\rm GDE}\big)$, respectively. The corresponding mock data in the selected ON and OFF region are $n_{\text{on},i} =b_{\text{on}_{ij},\rm CR} + b_{\text{on}_{ij},\rm GDE}$ and $n_{\text{off},i} =b_{\text{off}_{ij},\rm CR} + b_{\text{off}_{ij},\rm GDE}$, respectively \cite{Abdallah:2018qtu}. Therefore, the log-likelihood ratio is given by
\begin{equation}
\lambda = \frac{\mathcal{L_{\text{on/off}}}(\mu,\doublehat{\beta_1},\doublehat{\beta_2},\tau)}{\mathcal{L_{\text{on/off}}}(\widehat{\mu},\widehat{\beta_1},\widehat{\beta_2},\tau)} \,,
\label{eq:like2}
\end{equation}
where $\widehat{\mu}$, $\widehat{\beta_1}$ and $\widehat{\beta_2}$ are the unconditional ML estimators of $\mu$, $\beta_1$ and $\beta_2$, and $\doublehat{\beta_1}$, and $\doublehat{\beta_2}$ are the conditional ML estimators as a function of $\mu$. The calculated 5 sigma sensitivities from the ON/OFF analysis are presented as the black dotted and red dot-dashed curves in Fig.~\ref{fig:cta_onoff} for the low-cutoff and high-cutoff spectral models for the low-latitude FB data, respectively, which match with the models in certain energy ranges. Therefore, CTA will have good sensitivity to the FB edges. However, note that the selection of ON and OFF regions introduces trials in our ON/OFF study using Asimov dataset. Therefore the statistical significance would be reduced in future observations, when this look elsewhere effect~\cite{Lyons:1900zz} is taken into account.

To take into consideration of the systematics in ON/OFF analysis, first we compute $\theta_{\text{on},i}$ ($\theta_{\text{off},i}$) for the ON (OFF) region in each energy bin based on Eq.~(\ref{eq:sys}). The likelihood function is then formatted as below, 
\begin{equation}
\begin{split}
\mathcal{L_{\text{on/off}}}&= \prod_i \frac{ (\theta_{\text{on},i}m_{\text{on},i})^{n_{\text{on},i}}}{n_{\text{on},i}!}e^{-\theta_{\text{on},i}m_{\text{on},i}}e^{-\frac{(1-\theta_{\text{on},i})^2}{2\sigma^2}}\\
             &\times \frac{ (\theta_{\text{off},i}m_{\text{off},i})^{n_{\text{off},i}}}{n_{\text{off},i}!}e^{-\theta_{\text{off},i}m_{\text{off},i}}e^{-\frac{(1-\theta_{\text{off},i})^2}{2\sigma^2}} \, ,
\end{split}
\end{equation}
The obtained integral sensitivities with 3\% systematics can be seen in Fig.~\ref{fig:cta_onoff} as the upper dot-dashed and dotted curves for certain spectra. CTA sensitivity including this systematics is still good enough for FB detection of the high-cutoff model. However for the low-cutoff model, with limited number of energy bins in the analysis, the sensitivity for FB detection is only possible if systematics is $< 3\%$.

\subsection{Detection significance for LHAASO}
\begin{figure}[h!]
\begin{centering}
\includegraphics[width=3.15truein]{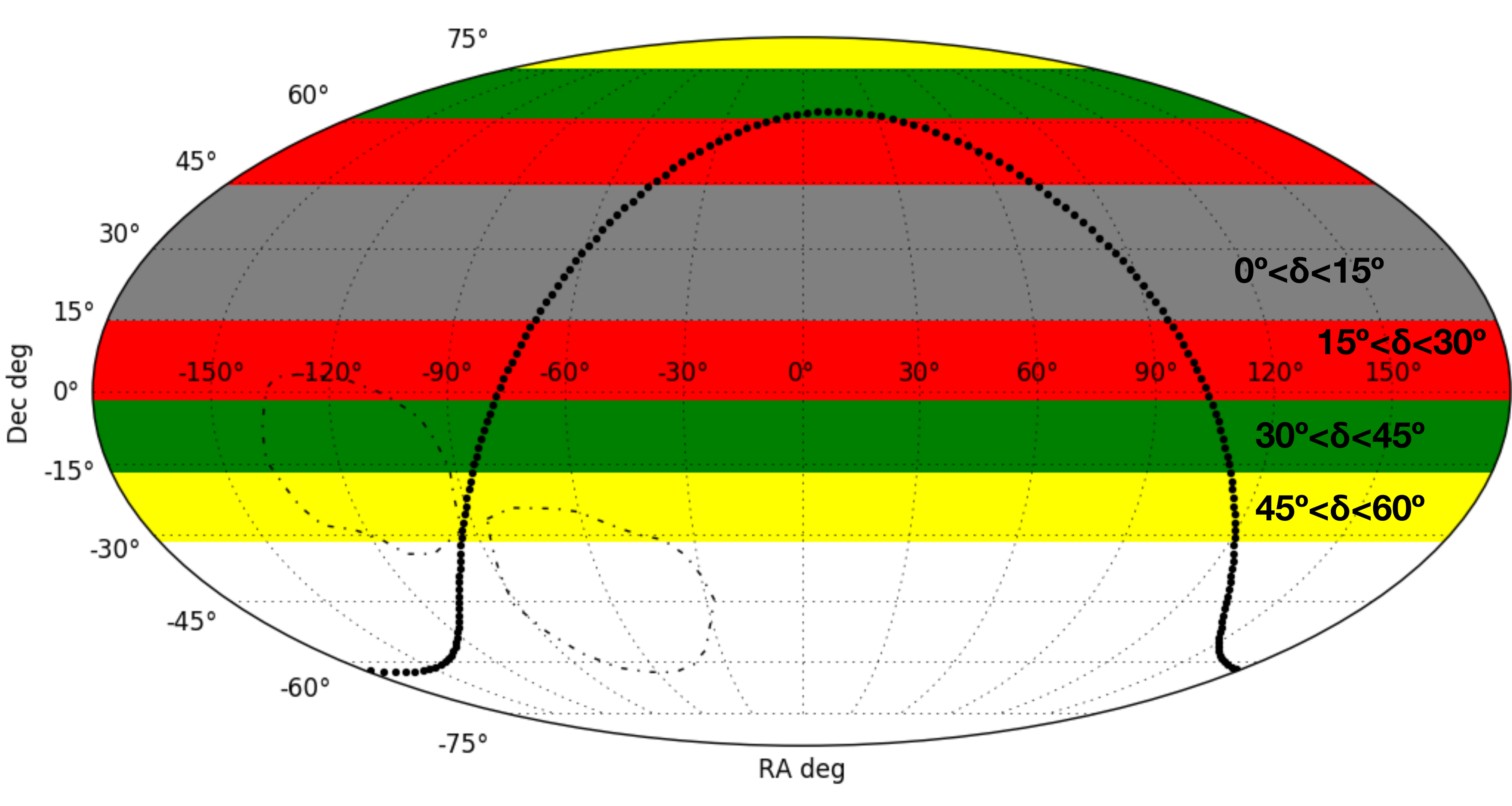}
\vskip -0.05truein
\caption{Field of view of LHAASO in 4 zenith-angle bands [0$^{\circ}$, 15$^{\circ}$], [15$^{\circ}$, 30$^{\circ}$], [30$^{\circ}$, 45$^{\circ}$] and [45$^{\circ}$, 60$^{\circ}$] as labeled in grey, red, green and yellow.  The bubbles are shown with dot-dashed contours. The black dotted curve represents the Galactic plane.}
\label{fig:lhaaso_fov}   
\end{centering}
 \end{figure}

LHAASO is located at a latitude of 29$^{\circ}$21'31" North and at a longitude of 100$^{\circ}$09'15" East. As one of the major components, WCDA has a coverage of 78,000~m$^2$ and observes gamma rays and cosmic rays in the energy range from 100 GeV to 100 TeV within a half-opening angle of 60$^{\circ}$.  The estimated scalar effective area \footnote{LHAASO Collaboration, private communication} can get up to 10$^6$ m$^2$ with energies above 30 TeV. According to its geographical location, LHAASO will be able to scan the whole Northern sky, and observe the Northern FB in 3 zenith-angle bands, [15$^{\circ}$, 30$^{\circ}$], [30$^{\circ}$, 45$^{\circ}$] and [45$^{\circ}$, 60$^{\circ}$] as shown in Fig.~\ref{fig:lhaaso_fov}.

As a ground based hybrid detector, the major concern for LHAASO is the gamma and hadron separation. Absorption of gamma rays due to electron-positron pair production with photons from starlight or cosmic microwave background is negligible in the tested energy range.  Another background is diffuse gamma-ray emission, which has been measured by \fl\ \cite{Ackermann:2012pya} and H.E.S.S.~\cite{Aharonian:2006au}.  To estimate the gamma-ray background in different sky regions, we adopt the same GDE model as CTA analysis from Ref.~\cite{Gaggero:2014xla} for the inner Galaxy region.  Since the cosmic electron spectrum is about two orders of magnitudes lower than that of hadron, its contribution to the background is neglected.  

We calculate the event numbers for FB, GDE and CR ($m_{\rm FB}$, $m_{\rm GDE}$ and $m_{\rm CR}$) for each zenith angle bin with formula, $m=T\int_{E_{\rm min}}^{E_{\rm max}}A(E) \phi(E)\Omega dE$, where $T$ is the exposure time and $\phi(E)$ is the flux of the FB, CR and GDE, respectively. Here, $A$ and $\Omega$ are the effective area and the daily averaged FB solid angle seen by LHAASO ($\sim 0.0003$, 0.026 and 0.07 sr) for three zenith-angle bins [15$^{\circ}$, 30$^{\circ}$], [30$^{\circ}$, 45$^{\circ}$] and [45$^{\circ}$, 60$^{\circ}$] separately. Based on the study of Monte Carlo simulations, the hadron/gamma separation is characterized by the quality factor $Q$~\cite{Proceedings:2017nxe}, where the hadron background can be reduced to $1/Q^2$ as expected. Hence, we obtain the model data $m_{kl}=\beta_1 m_{kl,\rm CR} + \beta_2 m_{kl,\rm GDE} + \mu m_{kl,\rm FB}$ and mock data as $n_{kl}=m_{kl,\rm CR}$+$m_{kl,\rm GDE}$ for 30 day observation, where $k$ and $l$ represent the $k$-th energy bin and $l$-th zenith bin. The likelihood function can be obtained as

\begin{equation}
\mathcal{L} =\prod_{k,l} \frac{m_{kl}^{n_{kl}}e^{-m_{kl}}}{n_{kl}!}.
\label{eq:lhaaso_likelihood}
\end{equation}

\begin{figure}[h!]
\begin{centering}
\includegraphics[width=3.15truein]{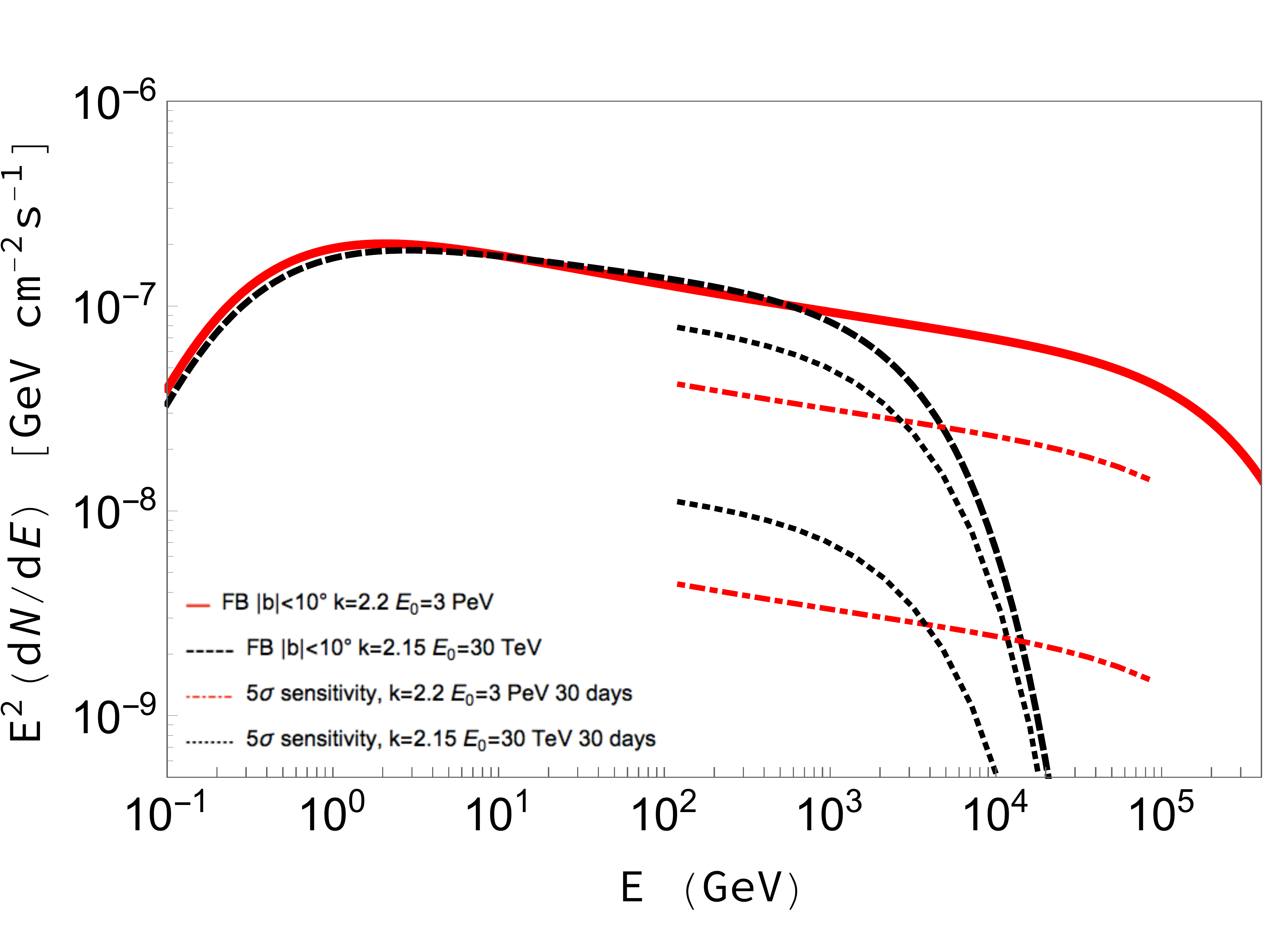}
\vskip -0.05truein
\caption{LHAASO 5 sigma sensitivity to the FB low-latitude high- (low-) cutoff flux model for 30-day observation with and without 10\% systematic uncertainty is presented in red dot-dashed (black dotted) curves in the energy between 100 GeV to 100 TeV respectively.}
\label{fig:lhaaso_sig}   
\end{centering}
 \end{figure}

We calculate integral sensitivity of LHAASO to the FB with 5 sigma significance for both the high-cutoff and low-cutoff spectra as seen in Fig.~\ref{fig:lhaaso_sig} with the same test statistical method as in Sec.\ IV A and IV B.  As a ground-base water Cherenkov detector, the overall systematic uncertainty, due to the detection efficiency,  air shower modeling, optical model and so on, is estimated to be $\sim$ 10\%~\cite{Abeysekara:2017old}.  In our calculation, we assume $\sigma = 0.1$ in Eq.~(\ref{eq:like1}) to take into account the systematic effect with the equations below, 
\begin{equation}
\mathcal{L}(n | m, \theta)=\prod_{k,l} \frac{(\theta_{kl} m_{kl})^{n_{kl}}}{\sqrt{2\pi}\sigma n_{kl}!}e^{-\theta_{kl} m_{kl}}e^{-\frac{(1-\theta_{kl})^2}{2\sigma^2}} \,.
\label{eq:like2}
\end{equation}
where
\begin{equation}
\theta_{kl}=\frac{1}{2}\Big(1-\sigma^2m_{kl}+\sqrt{1-2\sigma^2m_{kl}+4\sigma^2n_{kl}+\sigma^4m_{kl}^2}\Big) \,.
\end{equation}

Figure~\ref{fig:lhaaso_sig} shows LHAASO's sensitivity to the high- and low-cutoff spectra from the Northern bubble for 30-day observation as red dot-dashed and black dotted curves, respectively, where the upper (lower) curve is without (with) systematic errors. By including the systematic errors, our sensitivities are worsen by a factor of 7 to 10 for the high-/low- cutoff model. These estimations are for low-latitude bubbles, based on the same FB flux as for CTA. For CTA, we have enough instrumental information such as effective area as function of energy and zenith angle, PSF etc., and we bin our data in 3 dimensions accordingly. In other words we have sufficient statistic, so we calculate the differential sensitivity, as shown in Fig.~\ref{fig:FB_cta}. However, for LHAASO, we only have effective area for four zenith-angle bins, that means we do not have the spatial distribution of counts. So that is why in Fig.~\ref{fig:lhaaso_sig}, we present the integral sensitivity with and without systematic uncertainties and not differential sensitivity for each energy bin. With 10\% systematic uncertainty, LHAASO will be able to detect the Northern FB with 30 day of data in case of both high- and low-cutoff models.

\section{Summary and conclusions}
FB at the Galactic center \cite{Su:2010qj, Dobler:2009xz} are intriguing objects and are potential targets for ground-based VHE telescopes.  The analysis of Fermi-LAT data results in FB spectra being best described as having a cutoff above 100 GeV, in the high Galactic latitude ($|b| > 10^\circ$) region and without any apparent cutoff in the low Galactic latitude ($|b| < 10^\circ$) region \cite{TheFermi-LAT:2017vmf}.  The low-latitude hard spectrum, which is not constrained by recent HAWC observations \cite{Abeysekara:2017wzt}, can potentially be targeted by CTA and LHAASO in near future. 

We have explored sensitivity of CTA-South to the FB using Asimov data sets constructed from the FB hadronic flux models \cite{Razzaque:2018fzs} and by taking into account cosmic-ray and diffuse gamma-ray backgrounds.  We performed a morphological analysis, based on the proposed 525 hour Galactic center and 300 hour Northern Galactic plane observations by CTA \cite{Acharya:2017ttl}, using publicly available \texttt{ctools (v1.5.0)} software \cite{Knodlseder:2016nnv}. Our results show that CTA will be sensitive to the low-latitude FB in the $\sim$ 100 GeV --100 TeV ($\sim$ 100 GeV --7 TeV) range for high- (low-) cut-off spectrum.  A classic ON/OFF analysis with the same observations and flux models shows that CTA will be sensitive to the edges of the FB and put more constraints on our models. By taking into account 3\% independent systematic uncertainty in each pixel, the sensitivity is worsen by up to a factor of 10, especially in the energy range of 30--152 GeV due to the dominant CR background. At lower energy ($< 10$~TeV), the systematics determine the sensitivity, and at higher energy, the statistics take more control of the discovery. 
 
With less details known about the LHAASO observation strategies, we have performed a simple sensitivity calculation using expected events from the FB flux and background models, integrated in the 0.1--100 TeV range.  Our results show that LHAASO will have good sensitivity to the high-cutoff model for the low-latitude FB spectrum, even with a reasonably estimated 10\% systematics. However we also notice that the discovery of FB by LHAASO will crucially depend on the systematics and  sensitivity to the low-cutoff hadronic model will not improve even with longer observation time. Interesting upper limits can be derived from LHAASO observation in such a case. 

As planned, CTA will be completed by 2025 and LHAASO will be fully in operation until 2023. With 10 more years of exposure by then, the HAWC's sensitivity to the FB spectrum will improve, but still be above the high-latitude bubble. In other words, HAWC's sensitivity will not significant improve to constrain the FB spectrum. In future, a further analysis from HAWC will provide a better sensitivity, especially at lower energies and possibly larger search regions according to the predictions of some theoretical models.

The synergies between IACTs and EAS arrays \footnote{https://www.sgso-alliance.org/SGSOWiki/doku.php}, like CTA and LHAASO, one with very good sensitivity and angular resolution and the other with large field of view and high duty cycle, can be beneficial to observe the FB and either detect $\gtrsim 1$ TeV emission from the FB or constrain the gamma-ray flux models.  IceCube neutrino signal from the FB, if confirmed, will provide a chance to differentiate the leptonic and hadronic models through the observation of gamma rays and neutrinos. CTA will constrain the model parameters, such as the cut-off energy and spectral index. For example, with the detection of both messengers, the hadronic origin will be confirmed and the spectrum would be hard with high-cutoff energy. On the other hand, if the FB is observed by CTA at only at the sub-TeV energy range (overlapping with the \textit{Fermi}-LAT data), and without any neutrino excess, then the hadronic model with low-energy cutoff or a leptonic origin of the gamma rays will be preferred. The results will shed lights on the origin of FB and on multi-messenger (gamma ray and neutrino) emission from this intriguing source.   

\acknowledgements We thank Dmitry Malyshev, Cecilia Lunardini and Igor Oya for helpful discussion and for comments on the manuscript. We also thank Zhiguo Yao for providing  effective area data of LHAASO.  L.Y.\ acknowledges the start-up funding from SYSU and the Slovenia Research Agency grant number Z1-8139 (B). S.R.\ acknowledges support from the National Research Foundation (South Africa) with Grant No: 111749 (CPRR). 

This paper has gone through internal review by the CTA Consortium.
\bibliography{mybibfile}

\end{document}